\documentclass[12pt,a4paper]{article}       

\usepackage{xcolor}
 \usepackage[skins,theorems]{tcolorbox}
\tcbset{highlight math style={enhanced,
  colframe=red,colback=white,arc=0pt,boxrule=1pt}}
  \usepackage[bookmarksopen, bookmarksnumbered, bookmarksopenlevel=2]{hyperref}
 \usepackage[UKenglish]{babel}
 \usepackage[toc,page]{appendix}
 \usepackage{amsmath}
 \usepackage{amssymb}
 \usepackage{graphicx}
 \usepackage{hhline}

 \usepackage[bf]{caption}
\usepackage{cite}
\usepackage[vcentermath]{youngtab}
\usepackage{geometry}
\usepackage{slashed}
\usepackage{stackrel}
\usepackage{mathtools}
\usepackage{cancel} 
\usepackage{multirow}
\usepackage[margin=0pt,font=small,labelfont=normalfont,skip=22pt]{subcaption}

\usepackage{empheq}
\usepackage{arydshln}

\newcommand{\bqa}{\begin{eqnarray}}
\newcommand{\eqa}{\end{eqnarray}}

\newenvironment{eqn*}{\begin{equation*}\begin{aligned}}{\end{aligned}\end{equation*}\noindent}
\hypersetup{
    pdftitle={},
    pdfauthor={},
    pdfsubject={}
}
\numberwithin{equation}{section}
\numberwithin{table}{section}

\makeatletter

\definecolor{dgreen}{rgb}{0,0.45,0.2}
\definecolor{dblue}{rgb}{0,0.0,0.5}

\newcommand{\be}{\begin{equation}}
\newcommand{\ee}{\end{equation}}
\newcommand{\beq}{\begin{equation}}
\newcommand{\eeq}{\end{equation}}
\newcommand{\ba}{\begin{aligned}}
\newcommand{\ea}{\end{aligned}}

\newcommand{\bea}{\begin{eqnarray}}
\newcommand{\eea}{\end{eqnarray}}

\newcommand\bi{\begin{itemize}}
\newcommand\ei{\end{itemize}}

\def\unit{{1\kern-.65ex {\rm l}}}
\def\1{{1\kern-.65ex {\rm l}}}

\newcount\hour \newcount\minute
\hour=\time \divide \hour by 60
\minute=\time
\def\now{%
\ifnum \hour<13
  \ifnum \hour=0 \advance \hour by 12 \number\hour:\else \number\hour:\fi%
     \ifnum \minute<10 0\fi%
     \number\minute%
\ A.M.%
\else \advance \hour by -12 \number\hour:%
  \ifnum \minute<10 0\fi%
  \number\minute%
  \ P.M.%
\fi%
}

\makeatother

\begin{document}

\begin{titlepage}
\begin{center}
\rightline{\small }

\vskip 15 mm

{\large \bf
Swamplandish Unification of the Dark Sector
} 
\vskip 11 mm

 Cumrun Vafa

\vskip 11 mm

{\it Jefferson Physical Laboratory, Harvard University, Cambridge, MA 02138, USA}

\end{center}
\vskip 17mm

\begin{abstract}
We provide a short overview of recent progress made in our understanding of the dark sector based on the Swampland program which in turn is rooted in lessons from string theory.  We explain how the existence of one extra mesoscopic dimension (the ``dark dimension") in the micron range emerges and how this can lead to a unification of the dark energy and dark matter.
In particular the smallness of the dark energy leads to the prediction of the existence of a tower of weakly interacting light particles which can naturally play the role of dark matter.  Moreover this unifies dark matter with gravity as dark matter ends up being excitations of graviton in the dark dimension.  We also explain how in combination with other Swampland principles one finds an explanation of the ``why now" and the ``cosmological coincidence" problems.   This model is consistent with the cosmological bounds as well as the Newton's inverse square law, but makes predictions which differ from $\Lambda$CDM.  It also gives rise to an appealing picture of hierarchy of scales in particle physics pegged to the dark energy, including a possible origin of the electroweak hierarchy and the prediction of masses of QCD axion and sterile neutrinos both in the 1-10 meV range.  This review is intended for a broad audience of high energy theorists and cosmologists without prior knowledge of string theory and it explains the motivations and predictions of this program in a non-technical form.
\end{abstract}

\vfill
\end{titlepage}
\section{Introduction}

On the face of it the standard model of particle physics and the standard model of cosmology, the $\Lambda$CDM, have been very successful. However this hides the rather unnatural fine-tuning in the parameters of these models which have defied a simple explanation based on conventional approaches to naturalness.  For example on the particle physics side we have the hierarchy problem:  Why is the weak scale so small $\Lambda_W\ll M_{pl}$?  On the cosmology side, focusing on the late epoch, the smallness of the dark energy $\Lambda\sim 10^{-122}$ (in Planck units) and various coincidences are intriguing.  For example the {\it cosmological coincidence} problem:  why is it that when matter and radiation energy densities are equal they are also close to the dark energy density?\footnote{One explanation of this is the anthropic principle \cite{Weinberg:1987dv}: the dark energy is not much smaller than necessary for the formation of structure, which is needed for our existence.  The upper range of this value for dark energy leads to the explanation of this cosmological coincidence.  In this paper we offer a different explanation for this cosmological coincidence.} And the {\it why now} problem:  Why do we live in an epoch when the dark energy has just taken over $t_{now}\sim \frac{1}{\sqrt \Lambda}$?

The main approach to naturalness is typically rooted at symmetries which govern the physics at a given energy scale.  For example the IR physics would not care too much about the details of UV, except for conveying the symmetries at hand.  This UV/IR decoupling which has been central to the notion of naturalness seem to fail miserably for the issues mentioned above. And therefore hint at the need for a shift in the paradigm of how we go about describing the theory.

On the other hand in these considerations of naturalness, quantum gravity plays no role:  The naive belief is that gravity is strong at the Planck scale and by the time we get to the IR, it is irrelevant and so should not affect the content of the physical theory dramatically.  Precisely this point has been found to be incorrect based on what we have learned from string theory.  Indeed the fact that UV and IR cannot decouple in a quantum theory of gravity, should be expected:  For example if one considers big black holes, these can be viewed as objects whose existence follows from solving the Einstein equation in the IR.  On the other hand Bekenstein-Hawking entropy for black holes predicts that the degeneracy of the number of states at very high energy (UV), corresponding to the black hole mass, is given by ${\rm exp}(A/4)$ where $A$ is the area of the horizon of the black hole.  So the IR physics knows about the UV physics!
It is therefore not a strange thought to imagine the conventional approach to naturalness fails precisely because it ignores this UV/IR mixing.  In other words UV consistency of having quantum gravity in the mix places strong constraints on the IR physics. Thus the correct notion of naturalness should incorporate the constraints coming from the UV completion of quantum gravity, and this may ameliorate at least some of these fine tunings.

The question is how to find which effective field theories (EFT's) in the IR are consistent with quantum gravity in the UV.
String theory, which is a UV complete quantum theory of gravity offers a hope in this direction.  Indeed the solutions to string theory lead to very specific types of EFT's.  Moreover the example of string theory strongly suggests that almost all naively consistent EFT's are ruled out:  The set of consistent theories compared to naively allowed ones seems to be measure zero!  For example consider the maximally supersymmetric matter in 4 dimensions, ${\cal N}=4$ supersymmetric gauge theories coupled to ${\cal N}=4$ supergravity.  It can be shown, inspired by consistency conditions arising from string theory \cite{Kim:2019ths}, that the rank of the gauge group has an upper bound $rank(G)\leq 22$.  In particular $SU(N)$ gauge groups with $N\geq 24$ cannot be consistently coupled to ${\cal N}=4$ supergravity, despite the fact that they not only look naively consistent for all $N$, but in fact as quantum field theories decoupled from gravity are UV finite!

Given the scarcity of consistent theories of matter coupled to quantum gravity, it is easier to rule inconsistent theories out.  Indeed this has led to a program, the Swampland program, which summarizes the lessons we have learned from string theory into simple principles that can be used to assess whether the theories are allowed or are in the Swampland, i.e. naively consistent but ultimately inconsistent.  Even though we do not have a full list of consistency conditions, because we do not have a full understanding of string theory, we know enough about it to extract some fundamental principles which form the backbone of the Swampland program.  Many of these principles can also be interpreted and reinforced by other ideas of quantum gravity independently of string theory, and in particular from quantum aspects of black holes (see \cite{Agmon:2022thq} for a recent review of the Swampland program).  This cross checking with basic physics principles is necessary in order to make sure that we are not being blindsided by string theory examples, in case there exist other consistent quantum theories of gravity (even though this appears not to be the case).

The main aim of this short note is to explain the relevant parts of the Swampland program which have found surprisingly powerful applications in the context of the dark sector of our universe and in particular summarize the recent work \cite{Montero:2022prj,Gonzalo:2022jac,Law-Smith:2023czn,Obied:2023clp} leading to the prediction of the dark dimension and the unification of the dark sector. We also explain some consequences of the de Sitter conjectures \cite{Obied:2018sgi,Bedroya:2019snp} for the dark sector.  We also discuss predictions of this scenario for particle phenomenology, including an appealing picture of scale hierachy. The organization of this paper is as follows:  In section 2 we review aspects of the Swampland program relevant for the dark sector.  In section 3, we show how applying these principles to our universe leads to the prediction of one extra mesoscopic dimension in the micron range.  In section 4 we discuss the cosmological aspects of this model and explain how dark matter emerges in a natural way in this setup.  In section 5 we discuss experimental prospects for testing these ideas.  We conclude in section 6 with some observations about predictions of this scenario for particle phenomenology.

\section{Relevant Swampland Principles }
One of the main discoveries in string theory is the existence of duality symmetries.  If we take the parameters defining a physical quantum gravitational system to extreme values, a new `dual' description becomes more suitable.  The mechanism for this is that at extreme distances in field space, like large values of an expectation value of a scalar field $\phi$ we get a tower of light particles whose mass go down exponentially at large $\phi$ like 
$$m\sim {\rm exp}(-\alpha \phi)$$
for some $O(1)$ constant $\alpha$ (where all values are measured in Planck units). 
Based on string theory examples, it has been conjectured in \cite{Ooguri:2006in} that this is a general property of quantum gravity theories (the ``distance conjecture"). These light particles are weakly interacting and they provide the ingredients of the dual description.

\begin{figure}[ht]
    \centering
    \includegraphics[width=\textwidth]{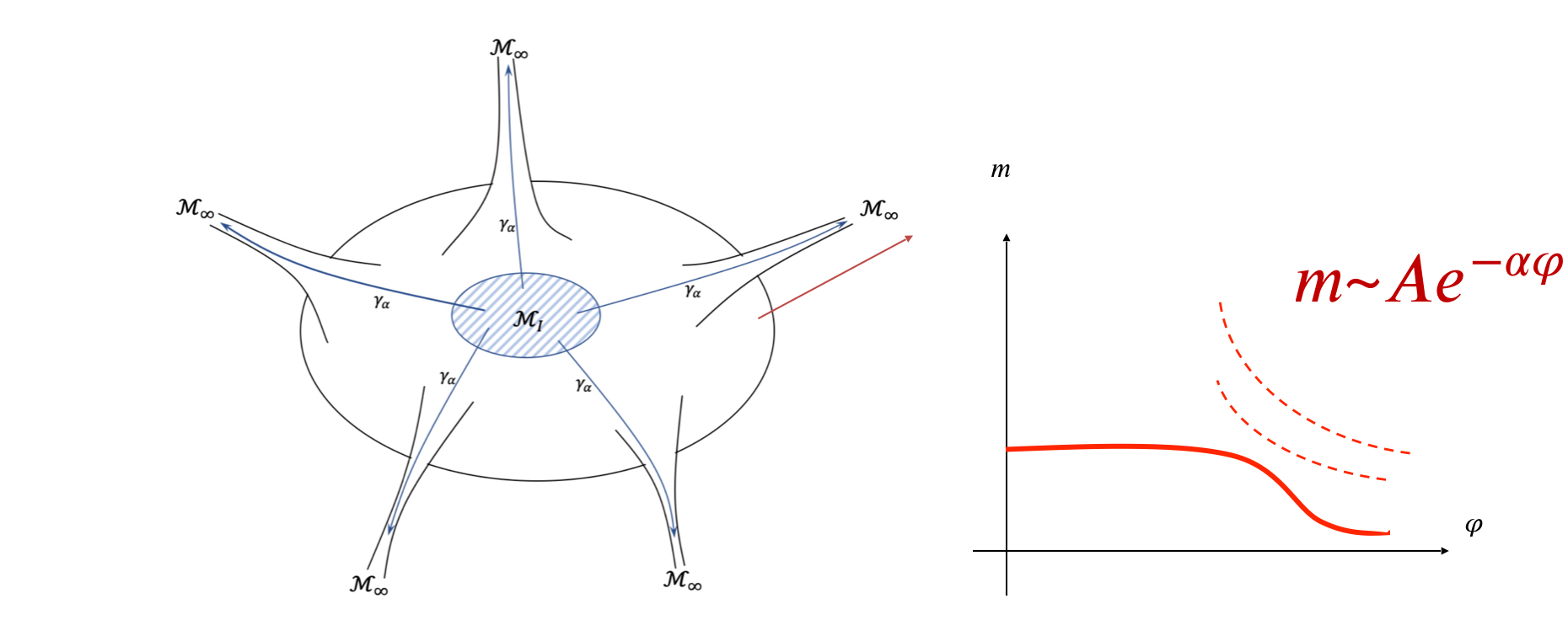}
    \caption{As we go to extreme values of field space, a tower of light particles emerge whose mass scales exponentially with field space distance.}
    \label{fig1:enter-label}
\end{figure}

  Moreover it is known that these light particles correspond to large wavelength excitations of graviton in higher dimensions (the KK modes), or a tower of excitations of a light string and it has been conjectured in \cite{Lee:2019wij} that this is a general property of all quantum theories of gravity (``emergent string conjecture").
The same reasoning applies to other parameters in the theory, for example the cosmological constant $\Lambda$.  In particular as $\Lambda\rightarrow 0$, since distance in metric space scales logarithmically with $\Lambda$, one gets a tower of light states which scale as \cite{Lust:2019zwm}
$$m\sim |\Lambda|^{a}$$
for some $O(1)$ constant $a$.  There is a lot of evidence for the latter statement for $\Lambda <0$ (in the context of AdS spaces). Even though it is very difficult to get a reliable $\Lambda >0$ de Sitter vacuum in string theory, the theoretical argument for the validity of the light tower applies equally well to positive or negative $\Lambda$.  Therefore it is natural to expect for it to be true for either sign.  For applications to our universe we would be interested in knowing the value of the exponent $a$ for $\Lambda>0$.  There is a heuristic argument that $a\geq 1/d$ \cite{Montero:2022prj}, where $d$ is the spacetime dimension.   The reason for this is that in the dual description, where the weakly interacting light tower are the basic ingredients of the theory, particles of mass $m$ in the tower contribute $m^d$ to the vacuum energy and thus the scale of cosmological constant cannot be smaller than that.  In supersymmetric theories the bosonic and fermionic contributions cancel but since here $\Lambda>0$ the theory cannot be supersymmetric and so they are not expected to cancel.  For $d=4$ this leads to $a\geq 1/4$ and the statement that 

\begin{equation}\label{bound}
m\lesssim \Lambda^{1/4}
\end{equation}
In the application to our universe we will assume this is a correct lower bound on the exponent $a$.\footnote{Regardless of this assumption, assuming a unification of dark energy and dark matter we explain in section 5 why in our universe $a> 0.2$, leading to a tower with mass scale less than $\sim 100 \ {\rm keV}$.}

Another Swampland conjecture, is the de Sitter conjecture \cite{Obied:2018sgi} or a more plausible version of it known as the ``trans-Planckian censorship conjecture (TCC)" \cite{Bedroya:2019snp}, which states that in an expanding universe regions smaller than Planck scale cannot exit the horizon and freeze, as sub-Planckian fluctuations are unphysical and should not become physically measurable.  This conjecture leads in particular to the prediction that potentials decay exponentially at large field values with exponents having a lower bound, 
\begin{equation}
    (V'/V)^2\geq \frac{4}{d-2}
\end{equation}
which has been verified in string theory constructions (see also \cite{Rudelius:2021oaz}).  A consequence of TCC is that even though it allows for the existence of a de Sitter space with dark energy $\Lambda$, it predicts that it cannot last longer than $\tau$ given by
\begin{equation}\label{tcc}
    {\rm exp}(\sqrt{\Lambda} \tau)\cdot 1 \lesssim \frac{1}{\sqrt \Lambda}\rightarrow \tau\lesssim \frac{1}{\sqrt \Lambda} \ {\rm log}(\frac{1}{\sqrt{\Lambda}})
\end{equation}
where we used the fact that the horizon scale is $1/\sqrt{\Lambda}$.  In other words up to logarithmic correction the maximum lifetime of a dS space is $\tau^{\max}\sim 1/{\sqrt \Lambda}$.

TCC offers a simple solution to the ``why now" problem, which is why do we live in an epoch which is of the same order as the dark energy horizon $t_{now}\sim \frac{1}{\sqrt \Lambda}$?  The explanation is that the maximum lifetime of dS according to TCC is of the order of the Hubble time $\frac{1}{\sqrt \Lambda}$, and therefore a typical time in such a universe before it decays is of that same order!

\section{Application to the dark sector of our universe}
We are now ready to apply these ideas to our universe.  The most spectacularly small non-zero parameter measured in fundamental physics is the dark energy $\Lambda \sim 10^{-122}$.  Thus it is natural to use this fact in the context of the Swampland conjectures to make concrete predictions for our universe.  The discussion below summarizes the results in \cite{Montero:2022prj}.

\subsection{Unification of the dark sector}
Using Eq. \ref{bound} we learn that there must be a tower of light KK particles or almost tensionless strings.
Without going into any further details we now explain why this implies a potential unification of dark sector:  the very existence of a small dark energy predicts the existence of a light tower of weakly interacting particles.
What can these weakly interacting particles be?  Why haven't we seen them in our universe?  The main point we will be explaining in this paper is that this light tower may be just the elusive dark matter we are looking for!  The first promising aspect of this is that we are predicting it to be very weakly interacting, just as is known to be the case for the dark matter.
Thus Swampland ideas leads to a unification of dark energy and dark matter:  The smallness of the former predicts the existence of the latter!

\subsection{The radius of mesoscopic dimensions}
To make this more specific we need to delve into more detail.  Eq.\ref{bound}
implies that the mass scale of the tower is less than $m\lesssim \Lambda^{1/4}\sim 10^{-2}\ {\rm eV}$.
However this light tower cannot be a string tower, because for light strings, gravity gets strongly coupled at that same mass scale.  This would have implied we would create black holes in scattering amplitudes in the eV regime, which is clearly ruled out by experiments.  So we learn that the light tower must come from some large extra dimensions, whose radius is bigger than or equal to
\begin{equation}\label{radius}
    R \gtrsim \Lambda^{-1/4}\sim 88 \ \mu m
\end{equation}
This in particular implies that Newton's inverse square law will be violated at the length scale smaller than $R$ where the force becomes stronger than expected:
$$F\sim 1/r^{2+n}$$
where $n$ is the number of larger extra dimensions.  
If \ref{radius} was viewed as a strict inequality it would be already ruled out because the inverse square law (ISL) experiments show that Newton's law is valid to about $30\ \mu m$ \cite{Lee:2020zjt,Tan:2020vpf}.  More precisely if the extra dimensions where flat tori, the experiments put this bound on the radius of circles, otherwise the experimental bound is on the smallest non-vanishing eigenvalue of the Laplacian for the internal compact space.  Thus for generic spaces there are order 1 numbers in the interpretation of the experimental bound.  Moreover the theoretical bound  \ref{radius} has also order 1 numbers associated with it.  Nevertheless the experimental result already implies that the mass of the tower must be in the eV -meV range and cannot be much bigger.  Thus the theoretical prediction combined with experimental observations has fixed the length scale of additional internal space to be in the micron range!

Finally we want to fix the number of extra dimensions $n$.  To explain what fixes this we need to explain more about how the standard model of particles can fit in such a picture, which we will now turn to.

\subsection{Standard Model on a brane}
In the context of larger extra dimension one can ask how the standard model fields fit?  If these fields lived in the bulk of the larger extra dimensions, they would also be accompanied by a tower of KK particles with the same quantum numbers. This would mean that for each particle in the SM we would get a tower of them separated in mass scale by meV-eV.  This clearly is contradictory with what we know from experiments.  There is no such tower with the same quantum numbers as the electron, for example.  Therefore the SM field cannot live in the bulk of extra dimensions.  However, there would be no contradiction if they are localized on a subspace of it.  This is very typical of how matter arises in string theory constructions, namely on a brane defect localized in extra dimensions.  Thus we end up with the only allowed alternative being that SM arises on a 3+1 dimensional brane localized in a (3+n)+1 dimensional space-time.  Note that the graviton propagates in the full space-time, including the $n$ extra mesoscopic dimensions.  In particular we have the coupling between SM fields and bulk gravitons:
\begin{equation}\label{bulkgraviton}
    \frac{1}{{\hat M}_p^{1+\frac{n}{2}}}\int d^4x \ T^{\mu\nu}_{SM}(x)h_{\mu\nu}(x,y=0)=\frac{1}{{\hat M}_p^{1+\frac{n}{2}}}\sum_n T^{\mu\nu}_{SM}(x) h_{\mu\nu}^n(x) \phi_n(0)
\end{equation}
where $y$ denotes the extra dimensions and $y=0$ is taken to be the position of the SM brane.  Here $\phi_n(y)$ denote the eigenfunctions of the Laplacian in the internal space, and $h_{\mu\nu}^n$ for $n\not =0$ denote the massive KK tower and $h_{\mu\nu}^0$ is the 4d massless graviton.
${\hat M}_p$ denotes the higher dimensional Planck mass, which is related to 4d Planck mass by
\begin{equation}\label{higherplanck}
    M_p^2={\hat M}_p^{n+2}V_n
\end{equation}
where $V_n$ denotes the volume of the internal n-dimensional space. Typically one expects, up to order 1 numbers $\phi_n(0)\sim {1/\sqrt{V_n}}$ and this leads to the expectation of an interaction of the form
\begin{equation}\label{finaleq}
\frac{1}{M_p}\sum_n \alpha_n \int d^4x T^{\mu\nu}_{SM}(x) h_{\mu\nu}^n(x)
\end{equation}
where $\alpha_n\sim O(1).$
This basically means that the standard model fields couple to each KK mode by gravitational strength, suppressed by the usual $1/M_p$.  Using this, one can estimate that if there is a thermal equilibrium on the SM at some temperature $T\gg m_{KK}$ the energy density leaks out to the bulk KK modes by the amount \cite{Arkani-Hamed:1998sfv}\footnote{The motivation for studying physical consequences of large extra dimensions in \cite{Arkani-Hamed:1998sfv} was to push the Planck scale down to the weak scale, which is not the case here, where the higher dimensional Planck scale is instead at $10^{10}\ {\rm GeV}$.}
\begin{equation}\label{leak}
    \frac{d\rho}{dt}\sim \frac{1}{\hat M_p^{n+2}}T^{7+n}\sim \frac{1}{M_p^2} T^{7+n}V_n
\end{equation}
Note that for temperatures larger than the inverse length scale of the internal manifold, the bigger the number of extra dimensions $n$ is, the more the energy leaks out to the bulk.

\subsection{How many extra mesoscopic dimensions?} 
So far we have argued, using a combination of Swampland ideas and experimental observations that there must be $n$ extra micron scale mesoscopic directions.  How big can $n$ be?  Using \ref{higherplanck} one sees that if $n>2$, then ${\hat M_p}<TeV$.  This is ruled out because in the colliders they would have led to large amount of production of black holes.  For the other two cases we have using \ref{higherplanck}
\begin{equation}
   {\hat M_p}\sim \Lambda^{\frac{n}
   {4(n+2)}}\Rightarrow  n=1 \rightarrow {\hat M_p}\sim \Lambda^{1/12} \sim10^{10}\ {\rm GeV} \qquad 
    n=2 \rightarrow {\hat M_{p}}\sim \Lambda^{1/8}\sim {\rm TeV}
\end{equation}
For $n=2$ the ${\hat M_p}$ is on the verge of being ruled out by collider physics as no black holes are observed in LHC. However $n=2$ can be ruled out by yet another experimental observation:  When neutron stars are formed by supernova events, they have a very high temperature.  Using \ref{leak} one can estimate how much KK modes are created during this process.  Some of these massive KK modes would be gravitationally bound to the neutron star and gradually decay back to pairs of photons, using the coupling \ref{finaleq}.  Absorption of some of this radiation by the neutron star continues to heat it up not letting the neutron star to cool beyond a certain amount. The observation of lower temperature neutron stars would put restrictions for such a scenario \cite{Hannestad:2003yd}.  For $n=2$ this puts a bound on the radius of the extra dimension to be $R<10^{-4} \ \mu m$, which is thus incompatible with the expected micron scale, again ruling out the $n=2$ extra dimensions.
For $n=1$ it turns out this bound leads to $R\lesssim 44\ \mu m$ which is just about the right range of allowed parameters! So the combination of the Swampland constraints and experimental observations lead to a unique prediction:  One extra mesoscopic dimension in the micron length scale:  the {\it dark dimension}!  In the next section we discuss how this scale which is motivated by the considerations of dark energy can also lead to the expected amount of dark matter in the form of KK tower of gravitons, leading to the unification of the dark sector with gravity.

\section{Cosmological aspects of the dark dimension scenario}
In this section we review a natural scenario for creation of dark matter in the dark dimension scenario \cite{Gonzalo:2022jac}.  For other cosmological approaches in the dark dimension scenario see \cite{Anchordoqui:2022txe,Anchordoqui:2022tgp,Anchordoqui:2022svl,Anchordoqui:2023etp,Anchordoqui:2024akj}.

As discussed in the previous section, we have been led by Swampland principles, combined with observations, to a unique corner of the quantum gravity landscape, with one extra mesoscopic dimension in the micron range.  Moreover a natural dark matter candidate in this scenario can be the KK excitations of the graviton in the dark dimension.  In this section we explain how this idea may work in more detail.

As far as evidence for the earliest time for thermal equilibrium in our universe, we believe that at the time Big Bang nucleosynthesis (BBN) took place the standard model matter fields were in thermal equilibrium at a temperature of $T_{BBN}\sim \ {\rm MeV}$.  Since the standard model matter is localized on a brane in the dark dimension scenario, this raises the question of whether the bulk 5d was also in thermal equilibrium with the SM fields?  The answer to this has to be no, because otherwise there would be many KK modes excited, which would be in contradiction with the observed $N_{eff}\sim 3$ (accounting for the light neutrinos only).  Therefore we conclude that the cosmology must include an epoch where only the SM brane is in equilibrium and the bulk is basically free of excitations.  Let $T_i$ denote the initial temperature where the SM brane was in thermal equilibrium.  We do not address the question of how it ended up there.
We also assume that the fields that set the mesoscopic dimension and the SM brane geometry are massive and are frozen to their fixed value.  In other words, we assume the universe behaves as in a 3+1 dimensional cosmology.

Starting from this initial temperature on the SM brane the usual big bang scenario will take hold, except for one major difference:  Since the SM fields couples universally to bulk gravitons it unavoidably creates them and some energy leaks to the dark dimension due to creation of KK gravitons. This leaking will lead to dark matter production in the form of KK gravitons.
Indeed from Eq. \ref{leak} we have (setting $M_p=1$ for simplicity of notation):
\begin{equation}
    \frac{d\rho_{DM}}{dt}\sim T^8\cdot R\sim T^8\Lambda^{-1/4}
\end{equation}
In the radiation dominated era $t\sim 1/T^2$
and so we can write the above equation in terms of $T$.
Let $y_{DM}=\rho_{DM}/T^3$ so it does not scale with the expansion.  Then the above equation (which ignores the effect of expansion) can be written as
\begin{equation}
        -\frac{dy_{DM}}{dT}\sim T^2\Lambda^{-1/4}
\end{equation} 
which we can integrate with respect to $T$ from $T_i$ to $0$ to find
\begin{equation}\label{ydm}
    y_{DM}\sim T_i^3 \Lambda^{-1/4}
\end{equation}
So the amount of dark matter produced is sensitive to the initial temperature we choose.  To get the right amount of dark matter we can simply fix the $T_i$ to the required value.  However, we will now try to motivate, based on Swampland principles, why the requisite $T_i$ is natural.

In order for the thermal equilibrium not to excite the internal geometry and the mass of the fields setting the internal geometry of the brane should be less than the initial temperature, i.e.
\begin{equation}\label{mass}
    T_i\lesssim m
\end{equation}
On the other hand it is natural to assume that at some earlier epoch these modes were excited and later settled to their final value.  However, this process could not have taken too long, because if we are to end up with a dS phase, according to Swampland ideas, it cannot last much more than the Hubble time $\Lambda^{-1/2}$.  In other words the rate of decay of these modes should be faster than or equal to
\begin{equation}
    \Gamma \gtrsim \Lambda^{1/2}
\end{equation}
On the other hand the moduli fields setting the internal brane geometry couple typically with gravitational strength amplitude $1/M_p$, so we expect their rate of decay to be $\Gamma \sim m^3/M_p^2$ and so we learn
\begin{equation}\label{bounding}
    m\gtrsim \Lambda^{1/6}
\end{equation}
We thus see that a natural solution to the bound \ref{mass} is to take
\begin{equation}
    T_i\lesssim \Lambda^{1/6}
\end{equation}
We will now show that for initial temperature in the upper limit of this bound, namely $T_i\sim \Lambda^{1/6}$ leads to the right abundance of dark matter! Indeed if we compute $y_{DM}$ using \ref{ydm} with this value of $T_i$ we learn that $y_{DM}\sim \Lambda^{1/4}$, which agrees with observations in our universe, where today we measure $y_{DM}=\rho /T^3 \sim \Lambda/\Lambda^{3/4}\sim \Lambda^{1/4}$. 

This scenario also sheds light on the ``cosmological coincidence" problem, i.e., why the matter/radiation equality temperature is close to the temperature where the dark energy takes over.
To see this let $y_{rad}=T^4/T^3=T$.  The matter radiation equality temperature $T_{MR}$(where the matter to first order is mainly the dark matter) is when $y_{rad}=y_{DM}$ which using \ref{ydm} leads to
\begin{equation}
T_{MR}=T_i^3\Lambda^{-1/4}\sim \Lambda^{3/6}\Lambda^{-1/4}\sim \Lambda^{1/4}=T_{DE}
\end{equation}
where $T_{DE}$ is the temperature at which the dark energy takes over.  So we find that with this initial temperature, not only we have the right abundance of dark matter, but also we have an explanation of the cosmological coincidence problem for any value of $\Lambda$!

\begin{figure}[ht]
    \centering
    \includegraphics[width=\textwidth]{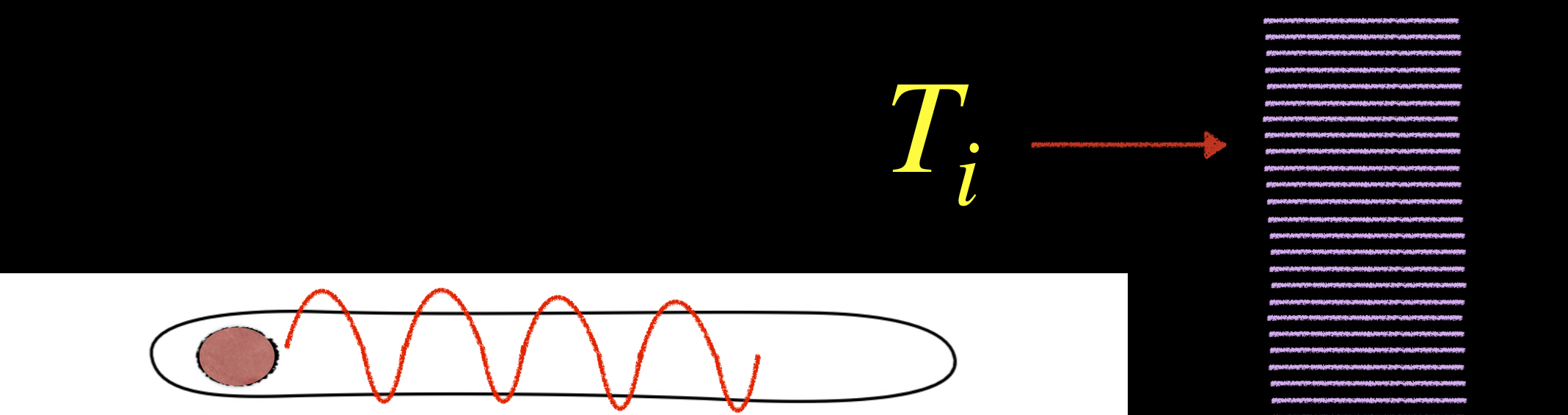}
    \caption{The standard model brane, which is localized in the micron scale 5-th dimension, radiates gravitons into bulk, which serve as KK dark matter.  Most of the initial dark matter distribution is at a mass close to the initial temperature.}
    \label{fig2:enter-label}
\end{figure}

Therefore we have a picture where we start the standard model brane at a temperature $T_i\sim \Lambda^{1/6}\sim {\rm GeV}$, which thankfully as is required is larger than the $T_{BBN}$, and the KK gravitons on the dark dimension are generated with enough abundance which at a later time, when the temperature is $T\sim \Lambda^{1/4}\sim eV$ will dominate the decreasing energy density of radiation and which is close to when the dark energy takes over.  The resulting dark matter composed of higher dimensional excitations of gravitons is weakly interacting (with gravitational strength) and so we end up with a model very similar to $\Lambda \rm{CDM}$, except for the following differences:  First of all the dark matter is not one species, but a tower of KK excitations.  Secondly the KK excitations are not stable and generically decay to lighter modes.  This realizes a particular version of the dynamical dark matter model proposed in \cite{Dienes:2011ja}.  Since there are strong bounds on the changing dark matter density, it must be that these decays to lighter elements of the tower do not lose much mass to kinetic energy.  A typical violation of KK quantum number $\delta$ cannot be too large and requires $\delta \sim O(1)$.
In the next section we discuss some experimental constraints for this dynamical dark matter model.

\section{Experimental constraints }
Here we summarize some of the experimental constraints on this model obtained in \cite{Gonzalo:2022jac,Law-Smith:2023czn,Obied:2023clp}.
Most of the dark matter is generated around $T_i\sim {\rm GeV}$ with this mass. Starting there they decay with gravitational strength to lighter elements of the tower given by $\Gamma\sim m^3/M_p^2$.

\begin{figure}[ht]
    \centering
    \includegraphics[width=\textwidth]{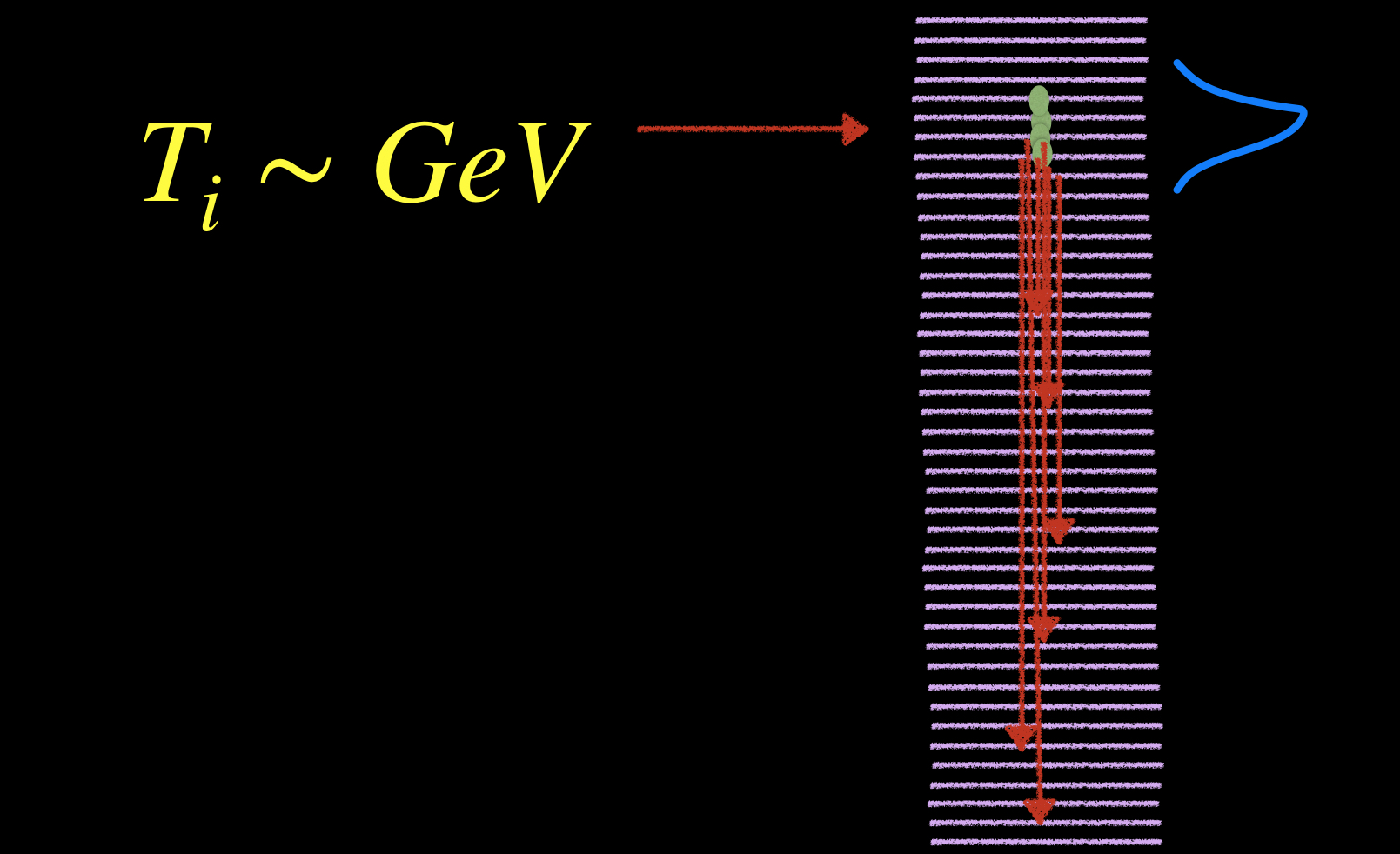}
    \caption{The KK gravitons are not stable and after they are created at a mass mostly around $\sim \rm{GeV}$ decay to lighter KK gravitons, today reaching a mass in the 50-100 $\rm{keV}$ range.}
    \label{fig3:enter-label}
\end{figure}

However there are many end states available.
There are $(m/m_{KK})^2$ possibilities for the pair of KK modes it can decay to.  However due to small violation of KK quantum number, this effectively becomes $(m/m_{KK})\cdot \delta$.  Moreover, taking into account the phase factor which is roughly the velocity of decay products at threshold $v\sim \sqrt{m_{KK}\delta/m}$ leads to
$$\Gamma^{tot}\sim
\frac{m^{7/2}\delta^{3/2}}{m_{KK}^{1/2}M_{p}^2 }$$
Setting $\Gamma \sim 1/t$ this leads to the prediction that the dark matter mass scales with time as
\begin{equation}
    m(t)\propto m(t_0) \big(\frac{t}{t_0}\big)^{-7/2}
\end{equation}

It turns out that this decrease in mass of the dark matter with time is crucial to avoid being in conflict with experimental observation.  In particular dark matter can decay back also to SM fields, and in particular to pairs of photons, with the rate $\Gamma\sim m^3/M_p^2$.  The higher the mass the more it decays to such photons.
It turns out that to avoid conflict with extra-galactic background radiation observations the mass of the dark matter today should be no larger than about $\sim \rm{MeV}$.  Moreover, the decay of KK gravitons to standard matter fields can distort the CMB and impact other astrophysical signals. Using this one can place bounds on the parameters of this model \cite{Law-Smith:2023czn}, and find that the natural range of $O(1)$ parameters in this scenario is consistent with these constraints leading to a mass of the dark matter today of the order of $\sim 100\ \rm{keV}$.  Putting $t_{now}\sim \Lambda^{-1/2}\sim 1/\Gamma^{tot}$ leads to the prediction that up to order one numbers in Planck units,
$$m^{DM}_{now}\sim \Lambda^{5/28}\sim 100\ \rm{keV}$$
the mass of the dark matter today is just barely within the limit allowed by observations thanks to its having decreased from $m\sim \Lambda^{1/6}\sim {\rm GeV}$ when it was originally produced.

Other predictions of this model of dark matter is that as it decays it gives a small kick velocity of the order $v\sim \sqrt{m_{kk} \delta/m}$ to the dark matter decay products.  This can impact the structure formation and observations puts a bound which again is of order one in the parameters of the model \cite{Obied:2023clp} leading to $v_{now}\lesssim 2\times 10^{-4}$.  Indeed 
\begin{equation}
v\sim \Lambda^{1/8}/\Lambda^{5/56}\sim \Lambda^{1/28}\sim 0.5\times 10^{-4}
\end{equation}
just within the allowed range.  It would be interesting to find imprints of this kick velocity in structure formation to distinguish it from $\Lambda {\rm CDM}$.

The most direct way to test the dark dimension scenario is to check Newton's gravitational inverse square law (ISL) at micron scale.  Due to $O(1)$ number ambiguities one can only predict this to appear at length scales $1-10 \mu m$.  Experiments checking this length scale would need to improve the current range bounds by a factor of 10.

\section{Concluding thoughts}

As we have seen in this review, Swampland principles combined with extreme smallness of the dark energy $\Lambda$ and other experimental observations lead to a unique promising corner of quantum gravity landscape with one extra mesoscopic dimension in the micron range, the dark dimension. This leads in particular to the unification of dark energy with dark matter realized as graviton excitations in this extra dimension.  We have also seen that various scales that appear in physics are powers of $\Lambda$:  
\begin{equation*}
    R^{meso}\sim \Lambda^{-1/4},\
    R^{macro}\sim \Lambda^{-1/2},\
     t_{now}\sim \Lambda^{-1/2},\
     {\hat M_p}\sim \Lambda^{1/12},\
    T_i\sim \Lambda^{1/6},\
    T_{now}\sim\Lambda^{1/4}
    \end{equation*}
    \begin{equation}
   m^{DM}_i\sim \Lambda^{1/6},\ m^{DM}_{now}\sim \Lambda^{5/28},\
    v^{DM}_{now}\sim \Lambda^{1/28}
\end{equation}
The fact that in this scenario many physical quantities are given by simple powers of the extremely small cosmological constant $\Lambda$ suggests that we may have found a realization of the hope expressed by Dirac \cite{Dirac:1979vf} for the explanation of appearance of large numbers in physics in terms of simple powers of one another. The explanation of electro-weak hierarchy may seem to be absent in the dark dimension scenario.  However, there has been a suggestion as how to also incorporate this as well \cite{Montero:2022prj}:  The neutrino mass $m_{\nu}\sim \Lambda^{1/4}$ can naturally be explained if we assume that the right-handed neutrinos propagate in the bulk 5-dimensional space as in large extra dimension scenarios \cite{Dienes:1998sb,Arkani-Hamed:1998wuz}, leading to the neutrino mass matrix
\begin{equation}\mathcal{M}\sim \left(\begin{array}{cc} 0&\frac{\langle H\rangle}{\sqrt{ R\hat{M_p}}}\\
 \frac{\langle H\rangle}{\sqrt{ R\hat{M_p}}}&\frac{1}{R}
\end{array}\right),\label{matrix}
\end{equation}
where $R\sim \Lambda^{-1/4}$.  This gives the right neutrino mass scale, assuming some couplings of order $10^{-2}$.
This will also lead to a tower of sterile neutrinos in the eV range.  If we assume that there is a mechanism making the neutrino sector non-hierarchic, i.e., that  the sterile neutrinos and the active neutrinos do not have vastly different mass scales, this leads to the condition that $$\frac{\langle H\rangle^2}{{\hat{M_p}R}}\sim (\Lambda^{1/4})^2\Rightarrow \langle H\rangle^2 \sim \hat{M_p}\Lambda^{1/4}$$
Using $\hat{M_p}\sim \Lambda^{1/12}$
this requires the weak scale to be
in the range $\Lambda_W\sim \Lambda^{1/6}\sim {\rm GeV}$, thus also potentially suggesting a new perspective on the electro-weak hierarchy!

This picture also leads to an interesting picture for QCD axion \cite{Gendler:2024gdo}: The axion naturally is localized on the standard model brane which leads to the prediction that the QCD axion decay constant $f_a\lesssim \hat{M_p}\sim 10^{10}\ GeV$.  Combined with observational lower bounds for QCD axion this narrows the axion mass to a small window in the range $1-10 \ meV$. Also $\hat{M_p}\sim \Lambda^{\frac{1}{12}} \sim 10^{10}\ GeV$ may signal a scale where new physics takes over. For example it could be an explanation of why the Higgs potential has an instability in this scale.

It is intriguing that
the neutrino, sterile neutrino, dark tower scale separation and the QCD axion all end up in the same range of about $\Lambda^{\frac{1}{4}}$.  Indeed in the dark dimension scenario the various scales in physics are all roughly organized in integral powers of the 5d Planck scale $\Lambda^{\frac{1}{12}}$:

\begin{align}
&\qquad \Lambda \sim 10^{-120} M_p^4 \notag \\
{\rm H}_0\sim \tau_{\text{now}}^{-1} & \sim \Lambda^{\frac{6}{12}}\sim 10^{-60} M_{p}\sim 10^{-40} \, \text{GeV} \notag \\
m_\nu \sim m_a & \sim \Lambda^{\frac{3}{12}}\sim 10^{-30} M_{p}\sim 10^{-10} \, \text{GeV} \notag \\
\Lambda_{\text{QCD}} \sim \alpha \Lambda_{\text{weak}} & \sim \Lambda^{\frac{2}{12}}\sim 10^{-20} M_{p}\sim \, 1 \,\text{GeV} \notag \\
\Lambda_{\text{Higgs inst.}} \sim {\hat M}_p\sim f_a & \sim \Lambda^{\frac{1}{12}}\sim 10^{-10} M_{p}\sim 10^{10} \, \text{GeV}
\end{align}

While based on all we have seen it is natural to expect that in the near future we will find exciting evidence for the dark dimension scenario, it is important to ask what would it mean, theoretically, if the experiments looking for deviations for ISL fail to observe it in the micron range?  Here we assess the certainty we have of the validity of various assumptions that go into predicting the dark dimension.
In the application of the Swampland's distance conjecture to dS it was argued based on naturalness that $m\lesssim \Lambda^{1/4}$ \cite{Montero:2022prj}.  If the exponent in this relation was different then the Swampland principle could still be valid but with a different exponent.  However, note that if we believe that the dark matter and dark energy are unified, as it is rather natural in the context of distance conjecture, using the fact that the  mass of the dark matter today cannot be smaller than about $100\ \rm{keV}$ (due to extra-galactic background radiation and distortions on CMB) leads to the prediction that $m_{KK}\leq 100\  {\rm keV}$, which implies $m\lesssim \Lambda^{0.2}$, and that the ISL experiments would need to find a deviation at scale no smaller than $10^{-4}\mu m$.  If indeed no deviation of ISL is found even at that scale, it could be that the dark matter and dark energy unification does not arise as expected and the exponent of the distant conjecture is even smaller than $0.2$.  Another option is that despite the massive amount of evidence for the distance conjecture for $\Lambda <0$, and the theoretical argument for it, which does not depend on the sign of $\Lambda$ perhaps there is a subtlety and it does not extend to $\Lambda >0$.  We can only wait for the experimental verdict.  Either way, we will learn exciting new physics!

\subsubsection*{Acknowledgments} 
I would like to thank my collaborators on the dark dimension scenario, whose joint works I have reviewed in this paper.  I would also like to thank Rashmish Mishra and Georges Obied for comments on this manuscript.

This work is supported by a grant from the Simons Foundation (602883,CV), the DellaPietra Foundation, and by the NSF grant PHY-2013858.

\bibliographystyle{jhep}
\bibliography{sample}

\end{document}